\documentclass[twocolumn,superscriptaddress]{revtex4}
\usepackage{graphicx}

\begin{document}
\title{Restoring of optical resonances in subwavelength hyperbolic etalons }

\author{A. Ciattoni}
\affiliation{Consiglio Nazionale delle Ricerche, CNR-SPIN, 67100 Coppito L'Aquila, Italy} \email{alessandro.ciattoni@aquila.infn.it}
\author{E. Spinozzi}
\affiliation{University of Rome ``La Sapienza'', Department of Information Engineering Electronics and Telecommunications, Via Eudossiana 18, 00184
Roma, Italy}

\begin{abstract}
We give a solution to the fundamental problem of restoring optical resonances in deep subwavelength structures by resorting to indefinite
metamaterials. We prove that a nanometric thick hyperbolic slab with very small permittivities exhibits etalon resonances and provides high-contrast
optical angular filtering. This is possible since the hyperbolic dispersion allows the vacuum radiation to couple with medium plane waves with
longitudinal wavenumbers large enough to yield optical standing waves within the nanometric slab thickness. Our findings can form the basis of a
novel way for shrinking optical devices down to the deep subwavelength scale.
\end{abstract}

\maketitle

Anisotropic metamaterials with permittivities of different signs (indefinite or hyperbolic media) \cite{Smit1,Elser,Nogin,WangW} both have an
intrinsic basic physics interest and are expected to play a fundamental role in the renewal of photonics mainly for the hyperbolic dispersion
characterizing their extraordinary plane waves. Basically, hyperbola inverted curvature and asymptotes existence are key ingredients providing
unusual and fascinating effects as, for example, negative refraction \cite{Smit2,Smit3,LiuL1,FangF,SunSu}, conversion of evanescent waves into
radiative waves \cite{LiuL2,YangY} and hyperlensing \cite{Jacob,LeeLe,Xiong,YaoYa,Casse}. Since ordinary waves through indefinite media still exhibit
standard dispersion, the extreme difference between the dynamics of ordinary and extraordinary waves  has been exploited to conceive optical devices
as polarization beam splitters \cite{Zhao1} and angular filters \cite{Aleks}. Recently, tunable indefinite media have been proposed and used for
devising efficient optical switches of subwavelength thickness where an externally applied electric field can literally rotate the extraordinary
waves hyperbola thus allowing to switch between radiative and evanescent waves \cite{Spino}. Besides, indefinite metamaterials have also proved to be
ideal platforms for creating optical models of relevant general relativity effects as the event of metric signature change \cite{Smoly}.

Spanning from quantum mechanics to optics, standing waves play a major role in physics since they provide a complete description of the
noninteracting system dynamics. Besides, standing waves are responsible for the resonances a system exhibits when it is externally perturbed by a
probe containing spectral components matching the allowed eigenfrequencies. In this work we face the fundamental physical problem of restoring
resonances in a structure with spatial size smaller than the external field wavelength, a phenomenology generally forbidden by the lack of standing
waves confined in a so small volume.
\begin{figure}[htb]
\centerline{\includegraphics[width=0.42\textwidth]{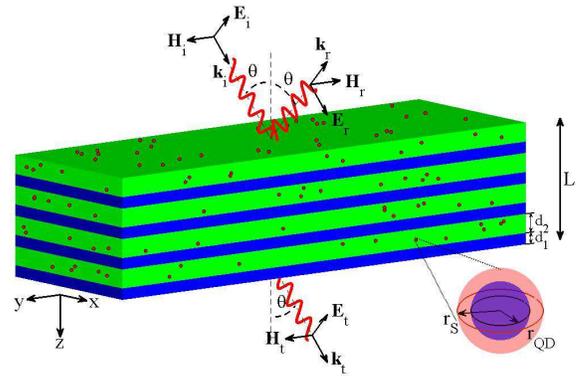}} \caption{(Color Online) Geometry of the metamaterial slab and of the inicident (i),
reflected (r) and transmitted (t) optical waves. The slab of subwavelength thickness $L$ is obtained by alternating nano-layers of an anisotropic
dielectric (1) and of a polymer hosting structured nanoparticles (2) whose internal structure, consisting of a core quantum dot (QD) and a metal
shell (S), is reported in the inset.}
\end{figure}
We show that a nanometric thick slab, despite its subwavelength size, displays very narrow optical etalon resonances if the medium is indefinite with
permittivities of amplitudes smaller than one. We point out that this is possible since the hyperbolic dispersion allows the vacuum radiation to
couple with medium plane waves of sufficiently high longitudinal wavenumbers to cause the formation of standing waves within the available
subwavelength slab thickness. Besides we prove that the restored resonances allow the slab to be regarded as a nanophotonic device able to provide
efficient angular filtering of optical radiation. Since the required electromagnetic properties are hardly exhibited by standard media, we propose
and design a realistic layered hyperbolic nano-composite and, numerically, we both check its electromagnetic homogenization and prove its
functionality as an high-contrast optical angular filter. We believe that, in addition to their conceptual value, our findings can have a crucial
importance for modern nanophotonics since they suggest a novel way for achieving optical functionalities at deep subwavelength scales.

Consider a slab of thickness $L$ illuminated by a monochromatic plane wave of wavelength $\lambda$ with its complex electric field ${\bf E}_i$
belonging to the incidence plane $xz$ ($p$-polarized or Transverse Magnetic wave) and impinging from vacuum with incidence angle $\theta$, as
depicted in Fig.1, so that its wave vector is ${\bf k}_i = k_0 \left(\sin \theta \hat{\bf e}_x + \cos \theta \hat{\bf e}_z \right)$, where $k_0 = 2
\pi / \lambda$. The slab medium is a uniaxially anisotropic homogeneous dielectric (slab layered structure of Fig.1 is used below) whose optical axis
is oriented along the $z$ direction so that the permittivity tensor is $\varepsilon = \textrm{diag} \left( \varepsilon_o, \varepsilon_o,
\varepsilon_e\right)$, where $\varepsilon_o$ and $\varepsilon_e$ are the ordinary and extraordinary permittivities. Since slab ordinary waves with
wave vectors in the $xz$ plane are polarized along the $y$-axis, in the considered interaction configuration only extraordinary plane waves are
excited with wave vectors ${\bf k} = k_x \hat{\bf e}_x + k_z \hat{\bf e}_z$ satisfying the dispersion relation $k_x^2 / \varepsilon_e + k_z^2 /
\varepsilon_o = k_0^2$. By enforcing the transverse momentum conservation across the slab interfaces, i.e. $k_x=k_0 \sin \theta$, we readily obtain
$k_z = k_0 \sqrt{ \varepsilon_o - (\varepsilon_o / \varepsilon_e) \sin^2\theta}$. The slab transmittance $T=|{\bf E}_t|^2 / |{\bf E}_i|^2$ (where
${\bf E}_t$ is the complex electric amplitude of the transmitted wave, see Fig.1) is easily evaluated after solving Maxwell equations and, if
absorption can be neglected (i.e. the permittivities are real), the familiar Airy expression
\begin{equation} \label{T}
T= \frac{1}{\displaystyle 1 + F^2 \sin^2(k_z L)}
\end{equation}
holds, where $F= 2k_z/(k_0 \varepsilon_o \cos \theta) +k_0 \varepsilon_o \cos \theta/(2k_z)$. From Eq.(\ref{T}) it is evident that the slab is fully
transparent (or $T=1$) if the resonance condition $k_zL=n\pi$ (where $n$ is any integer) holds, i.e. if the impinging wave couples to one of the slab
standing waves, this occurring for the incident angles
\begin{equation} \label{thn}
\theta_n = \arcsin \sqrt{ \varepsilon_e \left[ 1 - \frac{1}{\varepsilon_o} \left( n \frac{\lambda}{2L} \right)^2 \right]}.
\end{equation}
Standard etalons (for which $\varepsilon_o>1$ and $\varepsilon_e>1$) generally have thickness much larger than the wavelength (i.e. $\lambda / L \ll
1$) so that it is evident from Eq.(\ref{thn}) that a number of resonance angles exists, the square root being real and smaller than one for a number
of different integers $n$. Conversely, aimed at investigating possibly subwavelength thick etalons, we will assume that $\lambda / L \gtrsim 1$ so
that, if the principal permittivities assume standard values, etalon resonances are for all practical purposes forbidden since the square root is
imaginary for $n > 2\sqrt{\varepsilon_o} (L/\lambda)$. On the other hand if the permittivities have different signs (assuring the reality of the
square root of Eq.(\ref{thn})) and the extraordinary permittivity has a very small amplitude $|\varepsilon_e| \ll 1$ (yielding a square root smaller
than one for a number of different integers $n$), the considered subwavelength thick slab can exhibit various etalon resonances. From a physical
point of view, the mechanism supporting the resonance restoring can easily be grasped by inspecting the standing waves allowed by the slab.
\begin{figure}[htb]
\centerline{\includegraphics[width=0.45\textwidth]{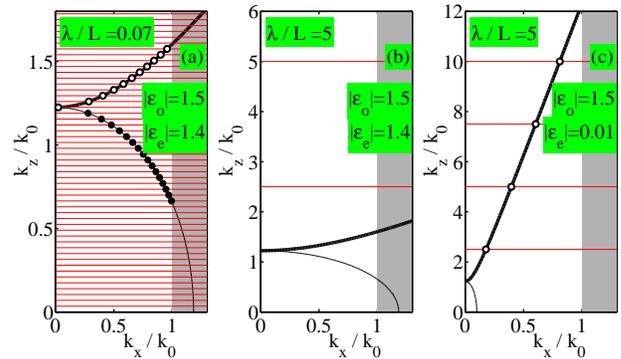}} \caption{(Color Online) Standard elliptic $k_x^2 / |\varepsilon_e| + k_z^2 /
|\varepsilon_o| = k_0^2$ (thin curves) and hyperbolic $-k_x^2 / |\varepsilon_e| + k_z^2 / |\varepsilon_o| = k_0^2$ (thick curves) extraordinary wave
dispersions together with the wave vectors $k_z = n \pi/L$ (horizontal lines) and mutual intersections corresponding to available slab standing waves
(closed and open circles). White stripes $k_x < k_0$ include waves which can be actually excited by the external plane wave.}
\end{figure}
In Fig.2 we plot the extraordinary plane waves dispersion relations both for standard elliptic $k_x^2 / |\varepsilon_e| + k_z^2 / |\varepsilon_o| =
k_0^2$ (thin curves) and hyperbolic $-k_x^2 / |\varepsilon_e| + k_z^2 / |\varepsilon_o| = k_0^2$ (thick curves) media together with the wave vectors
$k_z = n \pi/L$ (horizontal lines) and relative intersections (closed and open circles) corresponding to the slab standing waves. In Fig.2a
macroscopic etalons (i.e. $\lambda/L \ll 1$) are considered in the case case $|\varepsilon_o|>1$ and $|\varepsilon_e|>1$ and it is evident that a
large number of standing waves occurs both for standard and hyperbolic media. Conversely, as reported in Fig.2b, for subwavelength etalons (i.e.
$\lambda/L > 1$) characterized by $|\varepsilon_o|>1$ and $|\varepsilon_e|>1$, there is no available standing wave both for standard and hyperbolic
media since, for the formers, wave vectors $k_z = n \pi /L$ are too large (the thin line ellipse completely lying underneath the first horizontal
line) and, for the latters, standing waves cannot be excited by the externally impinging plane wave (the hyperbola being too widely opened to
intersect horizontal lines within the $k_x < k_0$ portion of the $k_xk_z$ plane). Resonance restoring in subwavelength etalons occurs for hyperbolic
media with $|\varepsilon_e| \ll 1$ since, as reported in Fig.2c, the hyperbola asymptotes are so close to the $k_z$ axis to allow the unbounded
hyperbola to intersect a number of horizontal lines within the white stripe $k_x < k_0$, thus yielding a number of standing waves which can be
actually excited through the plane wave impinging from vacuum.

If, in addition to the extraordinary permittivity, even the ordinary permittivity has a very small amplitude (i.e. $|\varepsilon_o| \ll 1$), it is
evident that $|F| \gg 1$ so that the transmittance angular profile of Eq.(\ref{T}) shows, around each $\theta_n$, approximately a lorentzian peak
whose width at half maximum is
\begin{equation} \label{dthn}
\Delta \theta_n = \frac{\displaystyle \frac{\lambda}{L} \left| \frac{\varepsilon_e}{\pi \sin \theta_n} \right|}
                       {\displaystyle 1-\left(\frac{L}{\lambda}\right)^2 \left(\frac{2\varepsilon_o}{n} \cos \theta_n \right)^2},
\end{equation}
as it is easily shown by using the expansion $F \sin(k_zL) \simeq \left[(-1)^n F L \left( dk_z/d\theta\right) \right]_{\theta_n} (\theta-\theta_n)$
valid for $\theta \simeq \theta_n$. It is remarkable that $\Delta \theta_n$ is here very small since $|\varepsilon_e| \ll 1$, as opposed to standard
etalons where the angular transmittance peaks are very narrow since $\lambda / L \ll 1$. In summary, a subwavelength thick slab exhibits etalon
resonances with associated narrow angular transmittance peaks only if the principal permittivities have different signs and have very small
amplitudes. Consider, as an example, an hyperbolic medium with principal permittivities $\varepsilon_o=0.2$ and $\varepsilon_e=-0.05$ which is
illuminated by a radiation of wavelength $\lambda = 829$ nm. In Fig.3 we plot the corresponding slab transmittance $T$ as a function of both the
incidence angle $\theta$ and the thickness $L$ spanning the subwavelength range $0<L<700$ nm and we note that the overall profile is characterized by
sharp ridges localized at the resonances curves $k_zL=n\pi$ (reported with dashed lines on the top of Fig.3). In addition, on the plane $L=700$ nm,
we have highlighted the angular transmittance profile with a bold line in order to emphasize its narrow peaks localized at the resonance angles of
the subwavelength-thick etalon.

An anisotropic medium with very small principal permittivities of different signs is hardly found in nature and therefore, in order to prove the
feasibility of the above discussed subwavelength thick device for optical angular filtering, we here numerically investigate one of its possible
realizations which we design through a suitable nano-composite. Indefinite dielectric permittivity is commonly achieved by resorting to
metal-dielectric layered structures \cite{Schil,Hoffm,Korob} so that, as reported in Fig.1, we here consider a slab obtained by alternating, along
the $z$-axis, nano-layers of an anisotropic medium $(1)$ and of an isotropic medium $(2)$, of thicknesses $d_1$ and $d_2$, with dielectric
permittivities $\varepsilon^{(1)} = \textrm{diag} \left( \varepsilon_o^{(1)}, \varepsilon_o^{(1)}, \varepsilon_e^{(1)} \right)$ and
$\varepsilon^{(2)}$, respectively. If the layers' thicknesses are much smaller than the radiation wavelength (i.e. $d_1+d_2 \ll \lambda $), the slab
exhibits the overall homogeneous optical response of a uniaxially anisotropic medium with optical axis lying along the stacking $z$-direction, i.e.
the effective dielectric permittivity tensor is $\varepsilon = \textrm{diag} (\varepsilon_o, \varepsilon_o, \varepsilon_e)$ where $\varepsilon_o =
f_1 \varepsilon_o^{(1)} + f_2 \varepsilon^{(2)}$ and $\varepsilon_e = \left[ f_1/\varepsilon_e^{(1)} + f_2/\varepsilon^{(2)} \right]^{-1}$ and $f_j =
d_j/(d_1+d_2)$ are the layers' filling fractions.
\begin{figure}[htb]
\centerline{\includegraphics[width=0.45\textwidth]{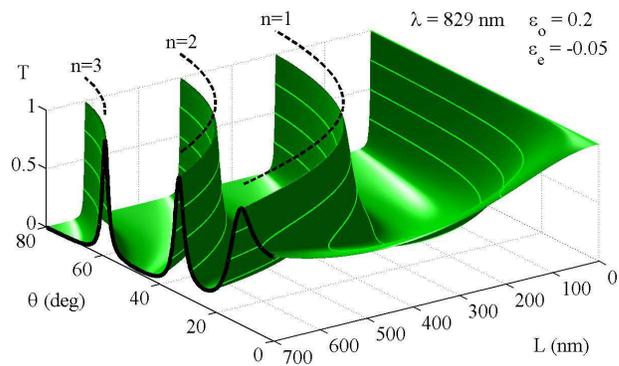}} \caption{(Color Online) Transimittance $T$ of an homogeneous hyperbolic slab as a
function of the incidence angle $\theta$ and the subwavelength thickness $L$. Dashed lines along the top plane and labelled with $n=1,2,3$ mark three
of the etalon resonance curves $k_z L=n\pi$ on the $\theta L$ plane.}
\end{figure}
In order for $\varepsilon_o$ and $\varepsilon_e$ to have different signs, we will require that medium $(1)$ is a transparent anisotropic crystal for
which $\varepsilon_o^{(1)}>0$, $\varepsilon_e^{(1)}>0$ and medium $(2)$ is such that $\varepsilon^{(2)}<0$. However, even though the requirement
$|\varepsilon_o| \ll 1$ is easily achieved by suitably choosing the filling fractions $f_j$, it is evident from the expression of $\varepsilon_e$
that the further requirement $|\varepsilon_e| \ll 1$ cannot be achieved if both $|\varepsilon_e^{(1)}|>1$ and $|\varepsilon^{(2)}|>1$. Therefore
medium $(2)$ cannot be a standard medium since $\varepsilon^{(2)}$ has to be real and $-1 < \varepsilon^{(2)} <0$. In order to achieve this goal, we
propose the layers of kind $(2)$ to be filled by a background transparent polymer of permittivity $\varepsilon_b$ hosting dispersed structured
nanoparticles $(NP)$ (see Fig.1) consisting of a core quantum dot $(QD)$ and a metal shell $(S)$ whose radii are $r_{QD}$ and $r_S$ (see the inset of
Fig.1) and whose permittivities are $\varepsilon_{QD}$ and $\varepsilon_S$, respectively.  The idea is that the metal shell provides the negative
dielectric constant whereas QDs provides the optical gain for compensating metal losses \cite{ZengZ}. Exploiting the electrostatic approach of
Ref.\cite{ZengZ}, the structured nano-particle can be described by the equivalent permittivity $\varepsilon_{NP} = \varepsilon_S [\varepsilon_{QD}
(1+2\rho) +2 \varepsilon_S (1-\rho)] / [\varepsilon_{QD} (1-\rho) + \varepsilon_S (2+\rho)]$, where $\rho=(r_{QD}/r_S)^3$, whereas the effective
permittivity of the whole medium $(2)$ can be evaluated through the standard Maxwell-Garnett mixing rule, i.e. $\varepsilon^{(2)} = \varepsilon_b
[\varepsilon_{NP} (1+2F) +2 \varepsilon_b (1-F)] / [\varepsilon_{NP} (1-F) + \varepsilon_b (2+F)]$ where $F$ is the nanoparticles volume filling
fraction. For a realistic material design, we choose typical II-VI semiconductor QDs (e.g. colloidal PbSe/ZnSe QDs) for which $\varepsilon_{QD}
(\omega) = \varepsilon^{(\infty)}_{QD} \left[ 1 + (6/\pi^2)(2A-1)\omega_{LT}/(\omega_n - \omega - i \gamma) \right]$ (where $A$ is the initial
occupation probability of the QDs ground state, $\varepsilon^{(\infty)}_{QD} = 12.8$, $\hbar \omega_n=1.5$ eV, $\hbar \omega_{LT}=5$ meV and $\hbar
\gamma=1$ meV) \cite{FuFuF} whereas we choose a silver metal shell for which $\varepsilon_{S}(\omega) = \varepsilon^{(\infty)}_{Ag} - \omega_p^2
/(\omega^2+i \Gamma \omega)$ (Drude-type silver permittivity with $\varepsilon^{(\infty)}_{Ag} = 4.56$, $\omega_p = 1.38 \cdot 10^{16} \:
\rm{s^{-1}}$ and $\Gamma = 0.1 \cdot 10^{15} \: \rm{s^{-1}}$). In order to exploit the optical gain provided by the QDs, we choose the wavelength
$\lambda = 829$ nm and $A=0.32$ so that $\varepsilon_{QD}=9.7804 - 0.6843i$ and $\varepsilon_{S}=-33.3299 + 1.66754i$. The electrostatic approach,
with the further choice $r_{QD}=8.2$ nm and $r_{S}= 9.9$ nm, yields $\varepsilon_{NP}=-3.3499 + 0.0006i$ and, after choosing $F=0.098$ and the PMMA
polymer matrix for which $\varepsilon_b =2.2020$, Maxwell-Garnett mixing rule finally yields $\varepsilon^{(2)} = -0.0468 + 0.0010i$, which has a
very small imaginary part and it is such that $-1 < \textrm{Re}(\varepsilon^{(2)}) <0$.
\begin{figure}[htb]
\centerline{\includegraphics[width=0.45\textwidth]{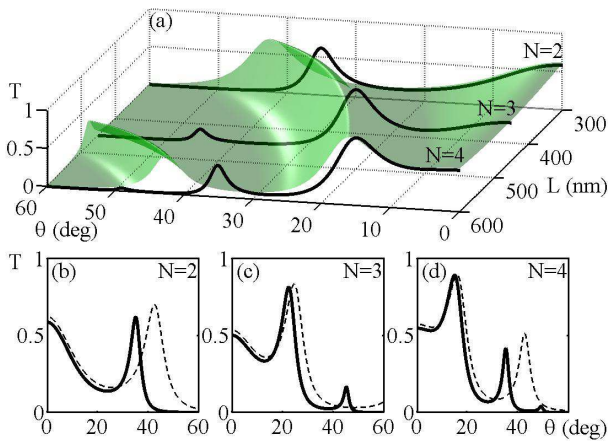}} \caption{(Color Online) Comparison between the slab transmittance $T$ evaluated for a
slab comprising $N$ bilayers with the transfer-matrix method (solid lines) and for the corresponding homogeneous effective slab (semi-transparent
surface of subplot (a) and dashed lines of subplots (b), (c) and (d)).}
\end{figure}
Besides, after choosing the layers of kind $(1)$ to be filled with alpha-iodic acid ($\alpha$-$\rm{HIO}_3$) with principal axis along the cartesian
axis (which can here be employed even though it is not a uniaxial crystal) for which $\varepsilon_o^{(1)} = \varepsilon_x^{(1)} = 3.3189$ and
$\varepsilon_e^{(1)} = \varepsilon_z^{(1)} = 3.8511$ and setting $d_1=10.5$ nm and $d_2=139.5$ nm for the layers' thicknesses, we finally obtain
$\varepsilon_o = 0.1887+0.0009i$ and $\varepsilon_e=-0.0504+0.0011i$ which are effective parameters satisfying the above discussed requirements for
observing the angular filtering slab functionality.

Since the chosen stacking period $d=d_1+d_2 = 150$ nm is not very much smaller than the adopted wavelength $\lambda = 829$ nm and layers of kind
$(2)$ are not strictly lossless, we have checked the above discussed nano-composite tailoring and the overall structure angular filtering
functionality by resorting to fully electromagnetic analysis. More precisely, using the standard transfer-matrix method, we have evaluated the
transmittance of various slabs comprising $N$ bilayers, by regarding layers of kind $(2)$ as homogeneous media with permittivity $\epsilon^{(2)}$. In
Fig.4 we plot the transmittance angular profiles $T(\theta)$ of various layered slab with $N=2,3,4$ bilayers (solid lines) and compare them with the
transmittances of the corresponding homogeneous slabs (semi-transparent surface of subplot (a) and dashed lines of subplots (b), (c) and (d)). Note
that the homogeneous approach correctly describes the transmittance for angles smaller than $\sim30^\circ$ whereas the agreement is qualitative for
higher angles and this is due to the fact that, the higher $\theta$, the smaller the medium radiation (longitudinal) wavelength $\Lambda = 2\pi / k_z
= \lambda / \sqrt{ \varepsilon_o - (\varepsilon_o / \varepsilon_e) \sin^2\theta}$ (for the considered effective parameters) and consequently the
worse the homogenization conditions. Also note that the transmittance angular peaks (predicted both by the layered and the homogeneous approaches) do
not reach the maximum value $1$ as a consequence of the residual absorption losses. However, despite the differences with its homogeneous counterpart
and the role played by absorption, it is evident that the considered nano-composite layered slab does actually show optical resonances, the ensuing
narrow peaks in the angular transmittance allowing to effectively regard it as a subwavelength thick angular filter.

In conclusion, we have shown that a slab whose thickness is smaller than the radiation wavelength supports standing waves and displays the associated
optical resonances if it is filled by an hyperbolic medium with very small principal permittivities. The transmittance angular profile exhibits
narrow peaks thus allowing the slab to be regarded as a nanophotonic angular filtering device. In view of the standing waves ubiquity and the
generality of the proposed method for restoring optical resonances in subwavelength structures, we believe that our findings might suggest a novel
way for devising nanometric optical structures with radiation steering functionality and consequently have a considerable impact on future
nanophotonics.

\end{document}